\documentclass[letterpaper, twocolumn]{article}
\usepackage{graphicx}
\usepackage{amssymb}
\usepackage[reqno,?]{amsmath}
\usepackage{bm, bbm}

\begin{document}

\newcommand{\pthz}[1]{
\mbox{$ #1\;\mathrm{pT}/\sqrt{\mathrm{Hz}}$}}

\newcommand{\fthz}[1]{
\mbox{$ #1\;\mathrm{fT}/\sqrt{\mathrm{Hz}}$}}

\onecolumn

\title{ \bf 
   All-optical magnetometry for NMR detection in a micro-Tesla field and unshielded environment}

\author {\it G.\,Bevilacqua, V.\,Biancalana, Y.\,Dancheva, L.\,Moi.}
\vskip 1 cm

\maketitle

\centerline {CNISM - Unit\`a di Siena, Dip. di Fisica and CSC Universit\`a di Siena,} 
\centerline { Via Roma 56, 53100 Siena, Italy.}

\begin{abstract}
An all-optical atomic magnetometer is used to detect a proton free-precession  signal from a water sample polarized in a
0.7\,T  field and remotely analyzed in a 4\,$\mu T$  field. Nuclear spins are manipulated  either by $\pi/2$ pulses or by non-adiabatic rotation. The  magnetometer operates at room temperature, in an unshielded environment and has a dual-channel sensor for differential measurements.

\end{abstract}

{\footnotesize{ \bf keywords}: Low Field NMR;  Optical magnetometry; Atomic magnetometry}

\section{Introduction}
\label{introduction}

The general trend in conventional nuclear magnetic resonance (NMR) spec\-tro\-scopy and magnetic re\-so\-nan\-ce imaging (MRI), has been to work at high fields, in order to increase the signal, and thus achieve better resolution with shorter measuring times. In fact, both the  magnetization and its precession frequency increase linearly with the strength of the magnetic field, so that the signal produced by Faraday induction in pick-up coils increases quadratically with it.

In fact, the polarization and the precession roles of the dc fields used in NMR can be discriminated. The two fields may differ in strength and/or orientation, and may be applied in two different regions (remote detection). 

When the time-dependent field produced by the precessing magnetization is detected by  non-inductive sensors, the signal amplitude no longer depends on the precession frequency. For this reason, methods based on non-inductive detection show their competitiveness  when the magnetization precesses in low magnetic fields. In this case the signal depends (linearly) only on the polarization field.

A recent  renewed interest in low-field NMR (LF-NMR) measurements has been motivated by the use of Superconducting Quantum Interference Devices (SQUIDs) as sensitive and frequency-independent magnetic flux detectors \cite{longquingjmr08, greenbergrmp98, mcdermottscience02, mcdermottpnas04}. Optical atomic magnetometers (OAMs) \cite{romalisnature07} are alternative sensors based on the magneto-optical properties of atomic samples.  OAMs were first proposed decades ago \cite{cohen69} and their performance has improved thanks to laser spectroscopy. Nowadays, OAMs achieve sensitivity levels comparable to those of SQUIDs  and are used in various fields of application \cite{leeaapl06, kornackprl05, belfijosa07, matsko05, weis03}, including  NMR detection \cite{savukovprl05a, belfijosa09, savukovjmr07, ledbetter09, savukovjmr09}.

The advantages of OAM sensors lie in the possibility of  miniaturizing the sensor volume \cite{schwindtapl04, ledbetterpnas08}, while  providing excellent time stability and reliability.  OAMs do not require cryogenics, as they work at room temperature or higher. This  feature brings a further advantage to  NMR as, besides dramatically reducing the cost of maintenance compared to SQUIDs, it helps to minimize the distance between sample and sensor, which is crucial for good sample-detector coupling.

From the point of view of sensitivity, SQUIDs operating at liquid He temperature reach a sensitivity in the  few \fthz{} range. At liquid N$_2$ temperature, this value increases to  tens of \fthz{}. These values are improved by a factor 30 in the case of high-Q resonator SQUIDs operating at several hundred kHz \cite{seton05}. For OAMs working in the so-called Spin-Exchange-Relaxation-Free (SERF) regime (which requires  magnetic field compensation down to fractions of nT), a sensitivity of \fthz{0.5} has been  demonstrated experimentally\cite{kominisnature03, savukovpra05}, while a fundamental limit of \fthz{0.01} has been claimed \cite{kominisnature03,  allredprl02}. For optical atomic set-ups working in a non-vanishing magnetic field the experimental limit is  \fthz{80}, with a theoretical projection as low as \fthz{1} \cite{xursi06}. Sensitivity as good as \fthz{1}  with projection as low as \fthz{0.01} have been  reported for OAMs working in non-vanishing fields, specifically designed to detect alternating magnetic fields and tuned to  resonantly oscillate with the time-dependent field to be measured \cite{savukovprl05b}.
One of the most significant differences  between SQUIDs and OAMs lies in the quantity measured. In fact, SQUIDs measure a component of the vector $\vec B$ while OAMs measure its modulus.

In LF-NMR experiments, the sample is typically first magnetized in an intense polarizing field, and then measured in a much weaker precession field. The second operation is  performed either after displacing the sample (remote detection experiments) or after switching off the polarizing field within a time interval much shorter than the relaxation time. Naturally, due to the high inductance of the coils producing the strong polarization field, such an abrupt field variation implies technical problems, and makes the second approach favourable only when remote detection is  not feasible (e.g. when investigating macroscopic solid samples). Furthermore, remote detection makes it possible to use permanent magnets for the polarization stage.

The  options available for spin manipulation change when  the strength of the precession field decreases. The field time derivative necessary to achieve non-adiabatic spin rotation decreases quadratically with the strength of the precession field, which makes it extremely easy to reach the non-adiabatic limit when working with precession fields in the micro-Tesla range. In fact, the transverse field used to rotate the spins, which has to be larger than the precession field, can nevertheless be much weaker than the polarization field, and the non-adiabaticity requires the transverse field to be switched off within a time interval shorter than the precession period (which is also inversely proportional to the precession field). 

As an alternative to using non-adiabatic rotation of the field, as in most conventional NMR experiments, a suitable time-dependent transverse field (the ordinary $\pi/2$ pulses) can be applied to rotate the magnetization with respect to the precession field. It is worth noting that in LF-NMR, all the spin manipulation pulses (e.g. the above mentioned $\pi/2$ pulses) must be at much lower frequencies: for this reason they must be referred to as ac pulses rather than rf pulses \cite{mcdermottscience02, yangapl06}. 
An obvious difference between the two spin manipulation approaches is  their nuclear selectivity. Should different nuclear species be studied at the same time, they  would all be reoriented with the first approach, while the resonant nature of the second approach would make the pulse  act selectively on a single species.
A discussion of the more advantageous procedures  for spin manipulation in LF-NMR for free induction decay (FID) detection by means of SQUIDs, can be found in \cite{longquingjmr08}. These procedures can, to some extent, be applied to LF-NMR with other non-inductive detectors.

We have previously demonstrated  that an OAM based on synchronous optical pumping of Cesium vapour is suitable for the detection of dc magnetization in prepolarized water \cite{belfijosa09}. In the present work, the same set-up is used to detect proton precession in water samples of a few cm$^3$ in volume. Here we report a set of experimental results, obtained using a LF-NMR setup  for remote detection of FID, with different  spin manipulation techniques and data-analysis approaches.   The experiment is performed with a permanent magnet polarization field in the 1T range, and a micro-Tesla precession field in which both the water sample and the OAM are immersed.

The OAM sensor works in an unshielded environment and has a differential nature, which makes compensation and shielding of stray magnetic fields less demanding in terms of accurateness. The set-up is relatively inexpensive and its operation is simple and largely automated. The long term stability (which is also improved by the automated control of the experimental parameters) partially compensates for the relatively poor sensitivity, as it permits noise rejection by long-lasting averaging.

Cheap and reliable set-ups for LF-NMR measurements  open up new fields of application  for NMR-based techniques, which have already demonstrated their potential in the construction of excellent diagnostic tools. Conventional NMR apparatuses, operating at relatively low fields are currently  used for the characterization of oil contents in bitumen \cite{byran08}, food analyses techniques \cite{Aeberhardt06}, and non-destructive on-line food quality control  \cite{andrade06}. In such  applications, LF-NMR can take advantage of the availability of large volumes with homogeneous precession field \cite{callaghanajp97}.

Most LF-NMR experiments reported in the literature refer to measurements performed in highly shielded volumes. Multiple layer shields of high permittivity material are used, guaranteeing excellent extinction of the environmental field. However, this approach prevent large samples from being analyzed, unless large (and thus very delicate and  expensive) shields are used. Testing LF-NMR performances in compensated but unshielded volumes is therefore clearly of interest for practical applications, such as developing  MRI set-ups for medical purposes.

\section{Set-up}
\label{Setup}

   \subsection{Overview}

The experiment  is performed as follows: water is pumped into a pipe and flows first into the high field region, where it is polarized, then into the weak field region, where spin precession is detected. Once there, the pump stops and data acquisition starts, lasting until the polarization decays. At this time the pump is switched on again and the whole cycle is repeated. 

\subsection{Water polarization system and water dynamics for remote detection}
Performing LF-NMR with a strong polarizing field makes it necessary to pass from the polarization to the precessing regime in a time interval that is shorter than the relaxation time. This can be difficult in both static and  remote sensing approaches. In the static approach, the problem lies in the large inductance of the polarization coils, while in the remote sensing approach the sample displacement must be large enough for it to escape from the spurious and inhomogeneous stray fields present in proximity of the prepolarization region.

The set-up contains a water polarization assembly comprising a set of permanent magnets. Soft iron shielding guarantees reasonable extinction of stray fields external to the polarization volume, and the
field/gradient compensation system described in Sec.\ref{compensation system} is sufficient to counteract the assembly's residual leakage. The polarization assembly comprises 56 cubic Nd magnets (1\,inch side) aligned in two rows of 70\,cm in length, placed at  7\,mm from each other, thus producing a field of
about 700\,mT. A pipe with a rectangular cross-section (30\,cm$^3$ in volume) is placed between the rows. The water is pumped into the pipe at a rate of  5\,cm$^3$/s, so it spends several seconds (several relaxation times) in
the polarization volume, and then flows through a capillary tube (1\,m in length and 1.8\,mm in
internal diameter) into the detection region for about 500\,ms. A bulb of 5\,cm$^3$ volume is placed close to the Cs cell of one arm, as shown in Fig.\ref{sensorandsample} (a). This bulb
contains a serpentine path to guarantee  efficient water refreshment, which is thus completed in less than 1.5~s.

    \subsection{Atomic magnetometer}
    \label{atomicmagnetometer}
All the NMR measurements discussed in this paper were performed with an  OAM based on the detection of non-linear Faraday rotation of light polarization, produced by optical pumping of Cesium vapour. The Cs OAM allowed us  to perform several kinds of high resolution magnetometric measurements \cite{belfijosa07}, including the detection of fields generated by cardiac currents \cite{belfijosacardio07} and the dc bulk magnetization of water \cite{belfijosa09}. Here we describe the main features of the magnetometer (more details are available in the papers cited above), and discuss its application in LF-NMR detection.

\begin{figure}
  \centerline{\includegraphics[width=7.5cm]{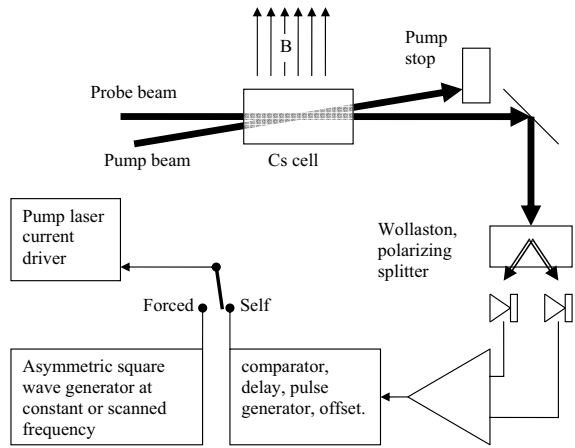}}
    \caption{Schematic of the magnetometer (main arm only).
          \label{setup1arm}}
\end{figure}

The OAM works with one or two channels, either in forced mode (with fixed or scanned forcing signal frequency) or as a self-oscillator. In the second case, the main arm gives the signal used to close the oscillator loop, while the second arm (if used) keeps working as a forced oscillator. A scheme of the main arm of the magnetometer is shown in Fig.\ref{setup1arm}.

Each arm contains a sealed glass cell containing Cs vapour and 90 Torr Ne as a buffer gas, which is heated to about 32$^\circ$C by means of circulating water, thus increasing the vapour density so that spin-exchange collisions become the  factor limiting the sensitivity \cite{savukovpra05}. The pump and the probe laser beams, which are are generated by independent single-mode free-running diode lasers, overlap within the cell. The lasers are passively stabilized by means of a high-quality current driver and temperature controller, and are both  tuned to the Cs $D_2$ line at 852\,nm. 
Laser sources with this wavelength are easy to find on the market thanks to their frequent use in telecommunications. It is worth noting that, as discussed in Sec.\ref{discussion}, relevant improvements could be made to the set-up performance if the pump laser were replaced by a source tuned to the $D_1$ line at 894\,nm.

While the probe laser delivers an unmodulated, linearly polarized beam,  the pump laser delivers a circularly polarized beam, whose optical frequency is made in resonant for short time intervals (pulses) via modulation of the laser junction current. At each pulse, the pump beam orients the atomic angular momentum along the propagation axis. 

Provided that the atoms are oriented synchronously with their precession (i.e. that the pulses occur at the atomic Larmor frequency  or 1/2, 1/3, ..., 1/N of its value), the medium shows a time dependent polarization, which precesses around the fields similarly to nuclear spins in NMR experiments, but with a frequency  two orders of magnitude higher. This precession produces a time-dependent rotation of the probe beam polarization plane, which is in turn detected by a balanced polarimeter composed of a Wollaston beam splitter and two photo-diodes. 
The polarimetric signal is thus synchronous with the pulses, and has maximum amplitude provided that the pulses occur at a frequency matching the atomic precession frequency (or integer fraction). 
The pulses may be generated by an external waveform-generator (forced mode) or triggered by the polarimetric signal, thus closing the loop and making the system self-oscillate.

The resonant frequency (or the frequency at which the system self-oscillates) varies proportionally to the magnetic field modulus through the atomic gyromagnetic factor, so that any change in the field modulus appears as a  modulation of the oscillation frequency.

In the presence of a dc bias field and a much smaller time-dependent field, the modulus changes proportionally to the vector component of the time-dependent field which is parallel to the bias field: 
$\mid \vec B \mid =
\mid \vec B_{dc} + \vec B_{ac} \mid=
( B_{dc}^2 +  B_{ac}^2 + 2  \vec B_{dc} \cdot \vec B_{ac} )^{1/2}
\approx B_{dc}+B_{ac} \cos{\theta}
$.

\begin{figure}
  \centerline{\includegraphics[width=7.5cm]{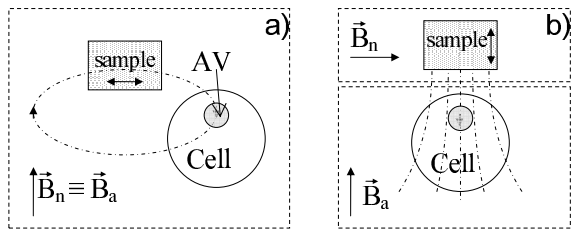}}
    \caption{Relative positions of the nuclear sample and atomic sensor. The nuclear magnetization precesses in the plane  (double arrow) perpendicular to $\vec B_n$, producing a time-dependent field (dot-dashed line) which must be parallel to $\vec   B_a$  in the active volume (AV) of a scalar sensor. This can be accomplished by
     a suitable angular displacement (case a) or (case b) by using $\vec B_n \perp \vec B_a$ .  \label{sensorandsample}}
\end{figure}

As sketched in Fig.\ref{sensorandsample}, detecting nuclear
precession with scalar sensors, makes necessary either to suitably displace  the nuclear sample with respect
to the sensor (case (a)), which is the configuration considered here, or to use
differently oriented fields for the atomic and nuclear precession
(case (b)). Solution (b) requires  homogeneous magnetic fields in two closely located regions, but  offers the 
possibility to adjust the nuclear and the atomic precession frequencies separately. As shown in \cite{savukovjmr07}, this  renders it possible to make the two precession frequencies resonant thus improving the detection sensitivity.

\subsection{Field and gradient compensation system} \label{compensation system}
The magnetic field measurement is based on the determination of the
frequency at which the atomic angular momentum precesses around the bias
magnetic field. As discussed below in Sec.\ref{sensitivity}, the uncertainty of this frequency depends on the noise level,  resonance slope and measuring time. Consequently, the sensitivity of OAMs is degraded by field inhomogeneities, which make the resonance broader and its central slope weaker. The set-up contains
large  coils to compensate and fix the three components of
the static magnetic field, and other electromagnets to control some components of the magnetic field gradient. The current supplies for field compensation are numerically controlled, thus enabling active recovery of the magnetic field drift.

As $B_z$ inhomogeneities ($z$ being the direction of the bias field)
are mainly responsible for the  reduction in sensitivity, in addition to three perpendicular Helmholtz coils pairs the set-up contains two pairs of dipoles. These are oriented along the direction of $z$ and placed on the $xy$ plane to control  the off-diagonal elements $\partial B_z /\partial x$ and $\partial B_z /\partial y$ of the gradient via a vanishing quadrupole field.

Additional smaller coils were used to introduce other oscillating or stepped fields  to manipulate the nuclear spin orientation. Oscillating fields resonating with the nuclear precession frequency and dc transverse field can be applied by means of coils with small inductance (a few tens of mHenry)  driven directly by a waveform or pulse generator. E.g. the transverse field can be switched off for a characteristic transient time L/R of 110\,$\mu$s.

\section{Experimental results} 

This section provides details of both the conventional spin manipulation techniques and describes the two approaches used for recording and analyzing the magnetometric signal. Experimental results  in the frequency and/or time domain are shown for both of the spin manipulation methods and both of the DAQ approaches.

\subsection{Spin manipulation}
\subsubsection{Manipulation with ac pulses}
\begin{figure}
  \centerline{\includegraphics[width=7.5cm]{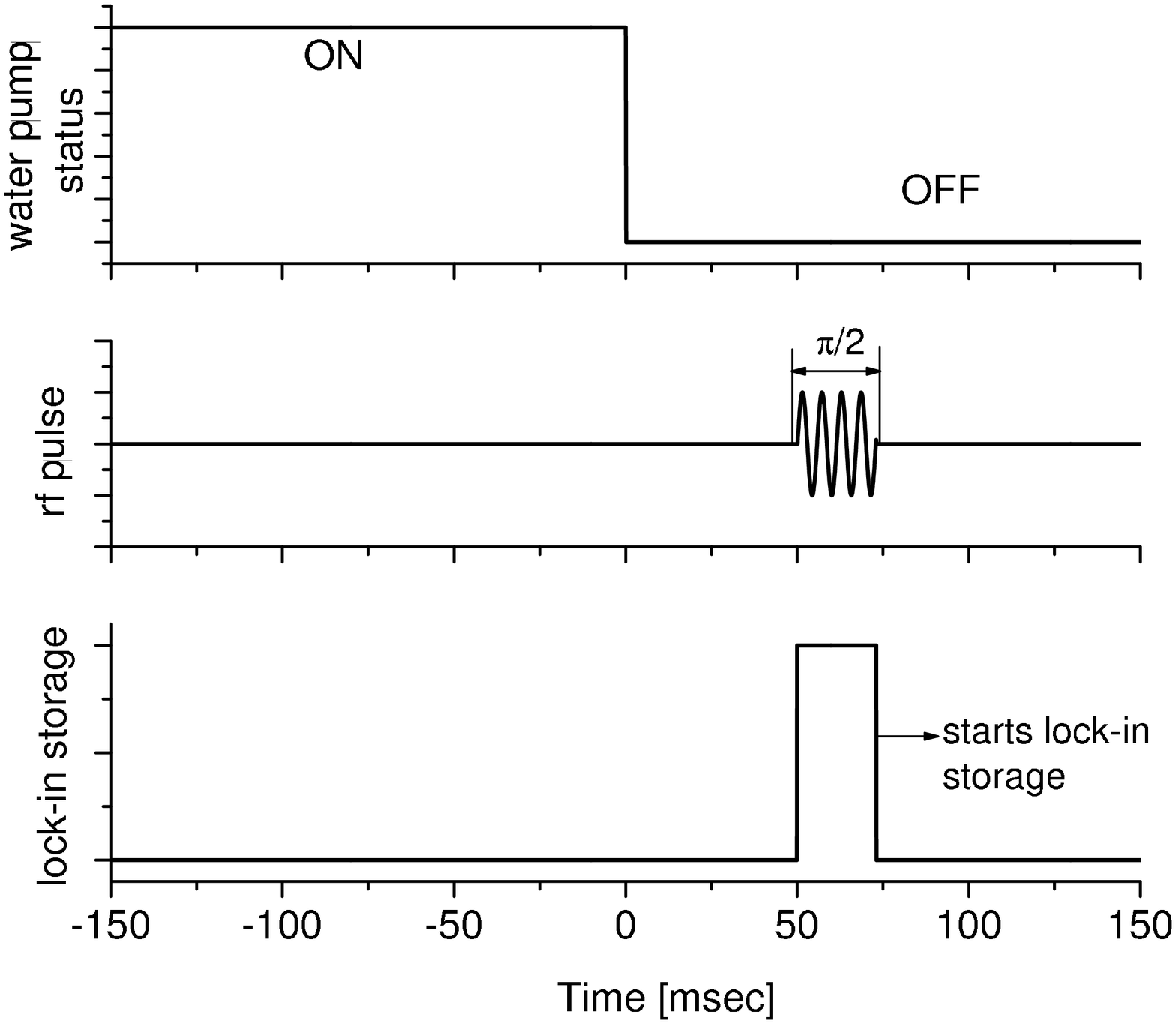}}
    \caption{Relative timing of water pump status, ac pulse and DAQ trigger. \label{time_seq_rf_pulse}}
\end{figure}

As mentioned above nuclear spins can be manipulated using pulses of transverse time-dependent magnetic fields in the in the LF-NMR a as in conventional NMR experiments. However as the nuclear spin precession  in this case is limited to few hundred Hz, supplying such pulse sequences is correspondingly simpler. In our set-up, a commercial waveform generator (Agilent 33250A) directly drives the current in the spin manipulation Helmholtz coils. A suitable user-defined waveform designed to generate the desired pulse sequences is uploaded and its timing is  controlled synchronously with pulses activating water flow and data acquisition. Fig.\ref{time_seq_rf_pulse} shows the relative timing of the ac $\pi/2$ pulse,  the water pump status and  the negative slope, which triggers the data acquisition system.

\subsubsection{Manipulation with non-adiabatic field re-orientation}
\begin{figure}
  \centerline{\includegraphics[width=7.5cm]{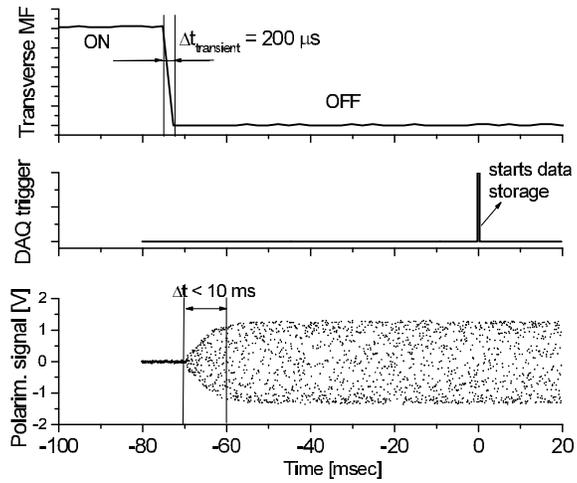}}
    \caption{Relative timing of the transition to {\it off} of the transverse precession field, growth of self-oscillation in the magnetometer  and DAQ trigger pulse. The {\it off-on} status of the transverse magnetic field has the same timing as  the water pump. \label{time_seq_non_adiab}}
\end{figure}

Manipulating nuclear spins by non-adiabatic field adjustments in LF-NMR is easy to achieve. In our set-up in addition to a  vertically oriented 4$\mu$T precession field, we have  a horizontal field of about 50$\mu$T which is
driven by a square-wave signal supplied directly by a waveform generator, via a 100\,Ohm resistor. This field has a transition time as short as $\delta t = 0.45$\,ms (corresponding to 4 times the L/R constant of the circuit), which perfectly fulfills the conditions for non-adiabaticity, $\omega_n \delta t \ll 1$, where $\omega_n / 2 \pi$ is the nuclear precession frequency.
Fig.\ref{time_seq_non_adiab} shows the timing of the transition of transverse precession field to {\it off}, the consequent growth of the self-oscillation in the OAM and the pulse used to trigger the data acquisition
system.

Nuclear spins enter  the transverse  field adiabatically  during the {\it on} period and flow into the measurement
bulb (sample magnetized in the $x-y$ plane). 
In fact, estimating   the  value of field gradient (both diagonal and off-diagonal components) as $B/D$, where $D$ is the typical distance from the field sources to the tube, the adiabatic transfer condition 
(small relative variation of $\omega_n$ for the displacement occurring in a precession period) reads:
$ (d \omega_n / dx) v_{flow}  \approx (\omega_n / D) v_{flow} \ll \omega_n/T$, $v_{flow} \ll D / T$. This condition is fulfilled in our case, as the water flow velocity is $v_{flow}\approx 2$\,m/s, $D\approx 10$\,cm, and $T < 6$\, ms.

\subsection{Data acquisition strategies}
\label{daq}

\begin{figure}
  \centerline{\includegraphics [width=7.5cm] {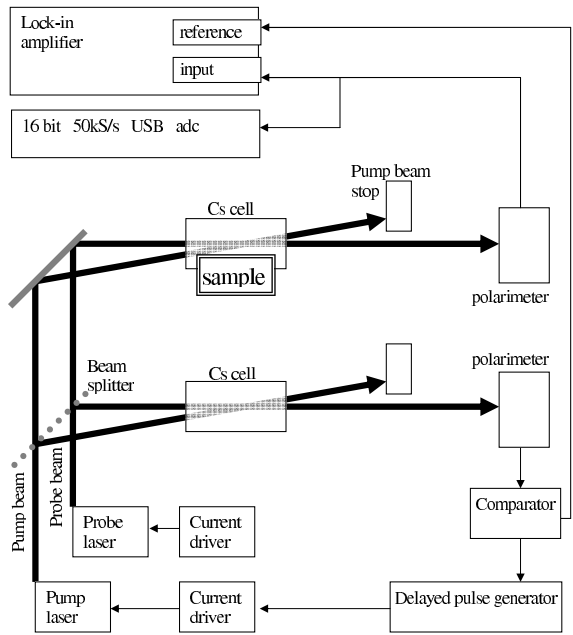}}
    \caption{Schematic of the differential set-up, details of the polarization devices and beam splitter are omitted. The lower (main) arm drives the self-oscillation at a frequency set by the environmental magnetic field, while the upper (slave) arm works as a forced oscillator, which may be slightly mistuned due to field variation caused by the sample. \label{setup2arms}}
\end{figure}

Two different approaches are developed for data acquisition and for extracting the NMR signal from the output signal produced by the OAM shown in Fig.\ref{setup2arms}.

 In the first approach,  the signal produced by the main arm of the magnetometer is directly digitized and numerically elaborated, while in the second approach signals from both arms (one  
locate close to the water sample and the other  sensing only the environmental field ) are used as input and reference signals for a digital lock-in amplifier (Stanford Research SR830). FID signals are reported, which were recorded using both of these approaches and  either ac pulses or non-adiabatic field rotation to cause the magnetization to precess.

\subsubsection{Direct digitization from a single channel}
Magnetization of the sample produces a variation of the magnetic field which, in the case of precessing
spins, is time-dependent. The OAM converts the instantaneous field modulus into atomic precession frequency, so a time-dependent field of modulus $B(t)=B_0+B_n(t)$ appears as an  instantaneous atomic Larmor frequency $\omega_a=\gamma_{a}  B(t)$. While $B_n(t)$ has a peak spectral component oscillating at $\omega_{L -
n}$, the bias field $B_0$  varies slowly in time due to random drifts in the environmental field, as well as to residual drifts in the  currents driving the compensation coils. In addition, due to the power net, $B_0$ contains large spectral components peaking at 50\,Hz and its harmonics.

The signal extracted from one polarimeter can thus be  modelled as:
\begin{equation}
V(t)=A \left[1 + \epsilon (t) \mathrm{e}^{i \phi_{N}(t) } \right]
\exp\left[i
\left(\omega_0 t + M \sin(\omega_{n}\, t) 
\right)
 \right]
\label{sigvstime}
\end{equation}
where $\omega_0$ is set by the dc component of the bias field. The
total noise $\zeta(t) = \epsilon(t)\mathrm{e}^{i \phi_{N}(t)}$ is
composed of the phase noise $\phi_{N}(t)$, which describes all
contributions  included the 50\,Hz noise produced by
the power net, and $\epsilon(t)$, which accounts for noise in the oscillation
amplitude.

\begin{figure}
  \centerline{\includegraphics[width=7.5cm]{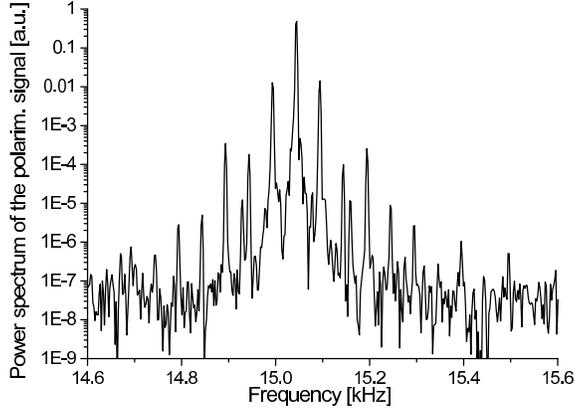}}
    \caption{ The power spectrum of the signal detected from one balanced polarimeter is shown. The central peak frequency is given by the atomic Larmor frequency, and sidebands clearly appear at 50\,Hz and multiples. In this single trace acquisition, the NMR signal produces sidebands well below the noise level. \label{pssingletrace}}
\end{figure}

Fig.\ref{pssingletrace} shows the power spectrum of the raw signal. The sidebands at 50\,Hz and multiples are clearly visible around the peak
at $\omega_0/2\pi=15.03$~kHz, while the sidebands corresponding to the NMR frequency, displaced by
$(\gamma_{n} / \gamma_{a}) \omega_0/2 \pi \approx 183$\,Hz are not distinguishable as they are the same as the random noise level or lower. The signal is acquired for finite time intervals, so that $\omega_0$ refers to
the given interval and may  vary slightly from one measurement to another. 
In order to extract the nuclear signal, the carrier angular frequency $\omega_0$ must first be evaluated precisely.
This  is done using a commercially available procedure, based on local analysis of the spectral peaks detected in the  power spectrum, evaluated by discrete Fourier transform of the Hann-windowed signal  \cite{lanczosgrandke}. The large signal-to-noise ratio seen in Fig.\ref{pssingletrace}  renders this procedure perfectly suited to the purpose.

The last exponential factor in Eq.\ref{sigvstime} is easy to  expand in Fourier series as

\begin{equation} 
\label{signalesp}
  \begin{split}
    \exp \left[ i \left( \omega_0 t + M\sin(\omega_{n}\,t) \right ) \right]  = &     
      \sum_{k=-\infty}^{\infty} 
      J_k(M)\exp \left[i \left(\omega_0+k\omega_{n}\right)t\right] \\
     & \approx
         \exp(i\omega_0t) \left [ J_0+ J_1 \exp (i\omega_{n} t)+
         J_{-1} \exp(-i\omega_{n} t) \right ] \\
     & \approx  \exp(i\omega_0t)
       \left [ 1 +  i M \sin ( \omega_{n} t) \right ], 
   \end{split} 
\end{equation}
where we have used the well-known properties of the Bes\-sel fun\-ctions:

\begin{equation} 
J_{-n}(z) = (-1)^n J_n(z), \mbox{and  } J_n(z) \approx (z/2)^n 
\end{equation}
valid for  $z \ll 1$.

Notice that the values of $\epsilon$ and $\phi_n$ are small, as can be inferred from the good contrast of the Fourier component at the carrier frequency with respect to  the noise, which exceeds  20\,dB for the 50\,Hz and 40\,dB for the broadband noise, as can be seen in Fig.\ref{pssingletrace}. According to Eq.\ref{signalesp}, in the spectral analysis  the nuclear signal appears  as a couple of peaks displaced by $\omega_n$ from the carrier frequency (sidebands),  which are hidden in the noise but contribute with a fixed phase with respect to the carrier.

After evaluating $\omega_0$ for each time interval corresponding to a measurement trace, we demodulate the signal by multiplying the digitized trace by $\cos(\omega_0 t)$. A discrete Fourier transform is then performed  using a FFT algorithm. The frequency axis is slightly rescaled on the basis of the observed $\omega_0$ drift, to make all the traces appear with superimposed peaks at $\omega_n$. The traces are finally  averaged, to filter out all the frequency components occurring with random phases with respect to the spin manipulation pulses.

\begin{figure}
  \centerline{\includegraphics[width=7.5cm]{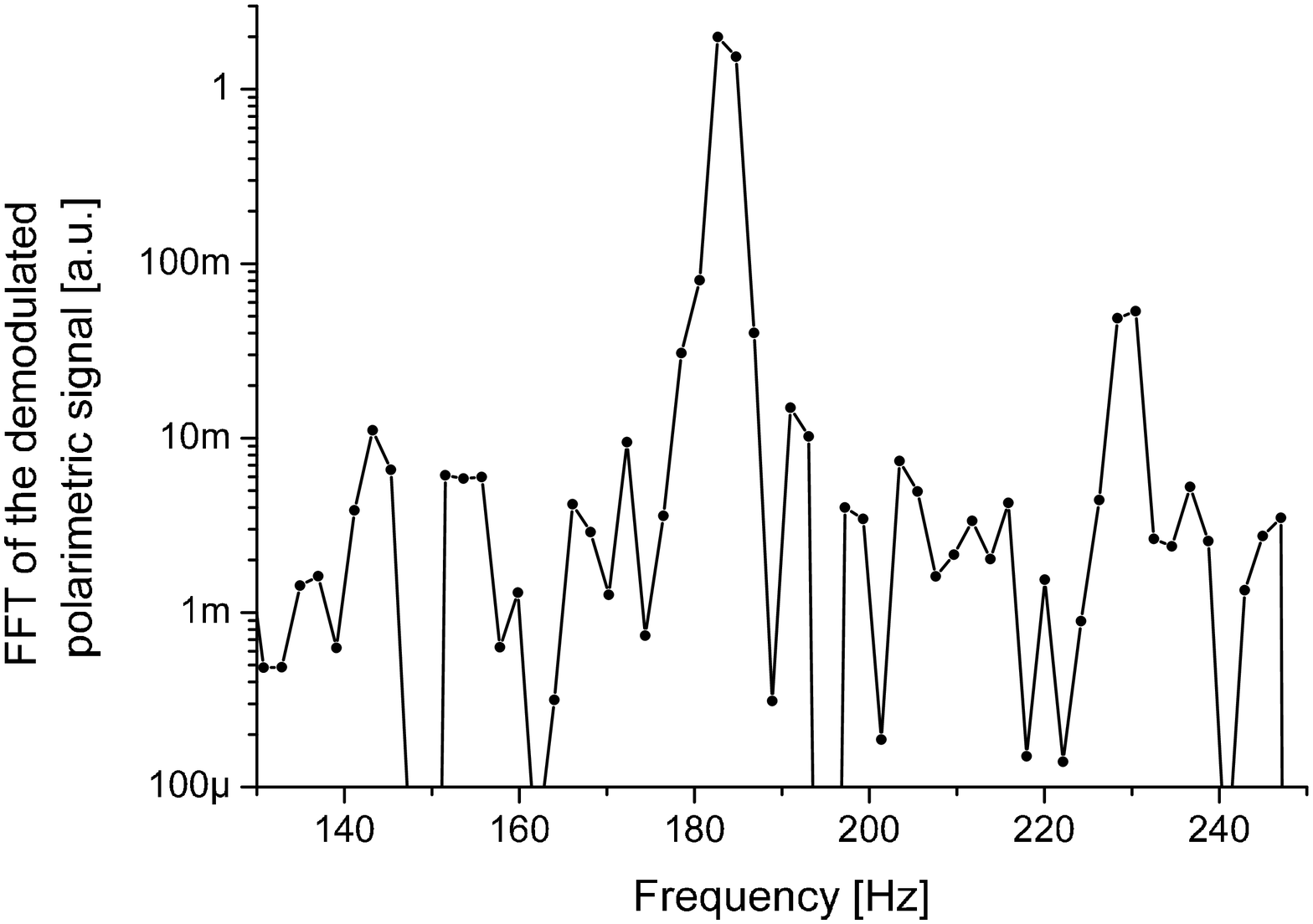}}
  \centerline{\includegraphics[width=7.5cm]{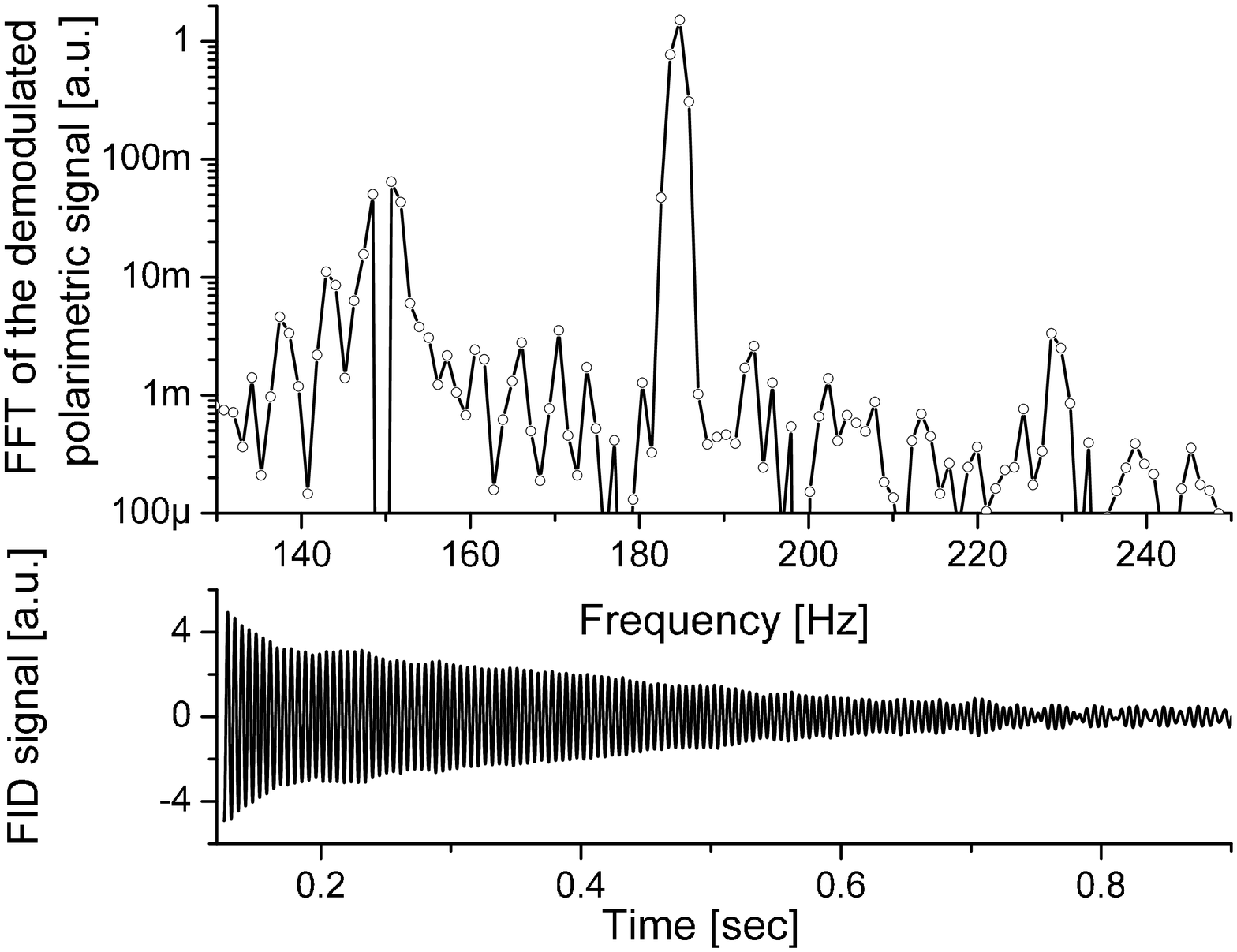}}
    \caption{Power spectrum of the FID signal obtained from the averages of 120, and 1350 directly digitized traces. The second signal is also shown in the time domain. These data are obtained by numerical demodulation of the sidebands produced by the time-dependent magnetic field. This field is generated by the nuclear FID responsible for frequency modulation of the atomic precession signal. \label{nmrsidebandtrace}}
\end{figure}

The same routine used to determine $\omega_0$  is also used to identify the amplitudes and the phases of the spurious peaks at 50\,Hz and its harmonics (typically up to the $4^{th}$), and to subtract those components from the demodulated trace. This cleaning procedure is essential when plotting the FID signal in the time domain. The time domain traces are filtered using a bandpass linear filter, in order to make   behaviour in time easily recognizable and the FID signal appreciable versus time on the average plot  (see Fig.\ref{nmrsidebandtrace}).

Our digitizing system is based on a 16 bit USB DAQ (MCC 1608 FS, Measurement Computing), which has a maximum sampling rate of 50\,kS/s and an on-card data buffer of 16383 points. This leads to a measuring time of 300\,ms,
when operating at the maximum sampling rate, which is too short with respect to FID time. The following  compromises may be considered: the sampling rate can be reduced to  34\,kS/s, increasing the measuring time to about half a second, and keeping the Nyquist frequency a couple of kHz above the atomic precession frequency $\omega_0/2 \pi \approx 15$~kHz; or (increasing in the noise level) one can work in undersampled conditions, so that  demodulation is performed using an aliased carrier peak. The FID trace shown in Fig.\ref{nmrsidebandtrace} was obtained by sampling at 18\,kS/s and demodulating with the alias of the 15\,kHz atomic carrier, appearing at 3\,kHz. Two frequency
domain plots, obtained from the averages of 120, and 1350 traces,  are shown together with the time-domain plot corresponding to the second signal.

\subsubsection{Lock-in detection (dual channel)}
Using a lock-in amplifier makes  it possible to take advantage of the differential nature of the set-up and to simplify  the data acquisition and processing. In this approach, the loop of the self-oscillating magnetometer
 closes on the arm  located farthest from the water sample, so that the instantaneous oscillation frequency tracks the environmental field. The corresponding signal is then used as a reference signal in a Phase Sensitive Detection (PSD)  system based on a digital lock-in amplifier. The signal of the polarimeter that analyses the light
emerging from the arm that senses the NMR, is applied to the lock-in input.

Let us assume that the time response of the lock-in amplifier is set to a time much longer  than the atomic precession period and much shorter than the nuclear one.  Its output can be considered as a measure of the steady-state oscillation amplitude of a forced and damped oscillator. The resonance is modelled with a Lorentzian curve  $A \Gamma /(\Delta\omega^2+\Gamma^2)$). The  signal amplitude can  be estimated from $A$, $\Gamma$, and from the instantaneous frequency deviation $\delta\omega_a$ of the atomic precession, which is in  turn caused by the field produced by the nuclear polarization. As the lock-in response depends approximately on the derivative of the resonance profile,  i.e. it is $A  /\Gamma^3$  in the Lorentzian case and at the resonance centre, the resulting expression for the lock-in output is
\begin{equation} 
V(t)=\frac{2A }{\Gamma^3} \delta\omega_a.
\label{lockinoutput}
\end{equation}

Two  limitations may apply to this approach. First, the re\-ference input of the lock-in amplifier must be capable  of tracking the reference instantaneous frequency. The reference lock-in input uses a PLL synthesizer whose filter specifications are not accessible to the user. We could only  verify that the time response was short enough to let the synthesizer follow frequency variations faster than we needed to cancel out the noise contribution overlapping (after demodulation) the NMR signal. 
Second, lock-in amplifiers are usually designed to generate slowly varying outputs and, although time
constants as short as 300~$\mu$s exist, a limited sampling rate is available for storing the lock-in output in the data buffer. In the case of our instrument, the maximum rate is 512\,S/s, setting  the Nyquist limit at 256\,Hz. It was operated at frequencies of about half of this limit to avoid aliasing effects, but the traces recorded  were roughly sampled. A data interpolating routine was implemented in order to achieve good visualization of the time traces and to perform slight frequency axis rescaling, which is necessary when averaging many traces in the presence of non-negligible drifts of the precessing field. The routine is based on direct DFT, zero padding, and inverse DFT (see. e.g. \cite{kau01}), possibly followed by appropriate re-sampling to permit superimposition of the frequency axes of subsequnent traces, as described above. As an alternative to resampling and rescaling, a numerical feedback with integral response can be used, adjusting the bias magnetic field  once per trace, to recover the drift of the atomic Larmor frequency.

\begin{figure}
  \centerline{\includegraphics[width=7.5cm]{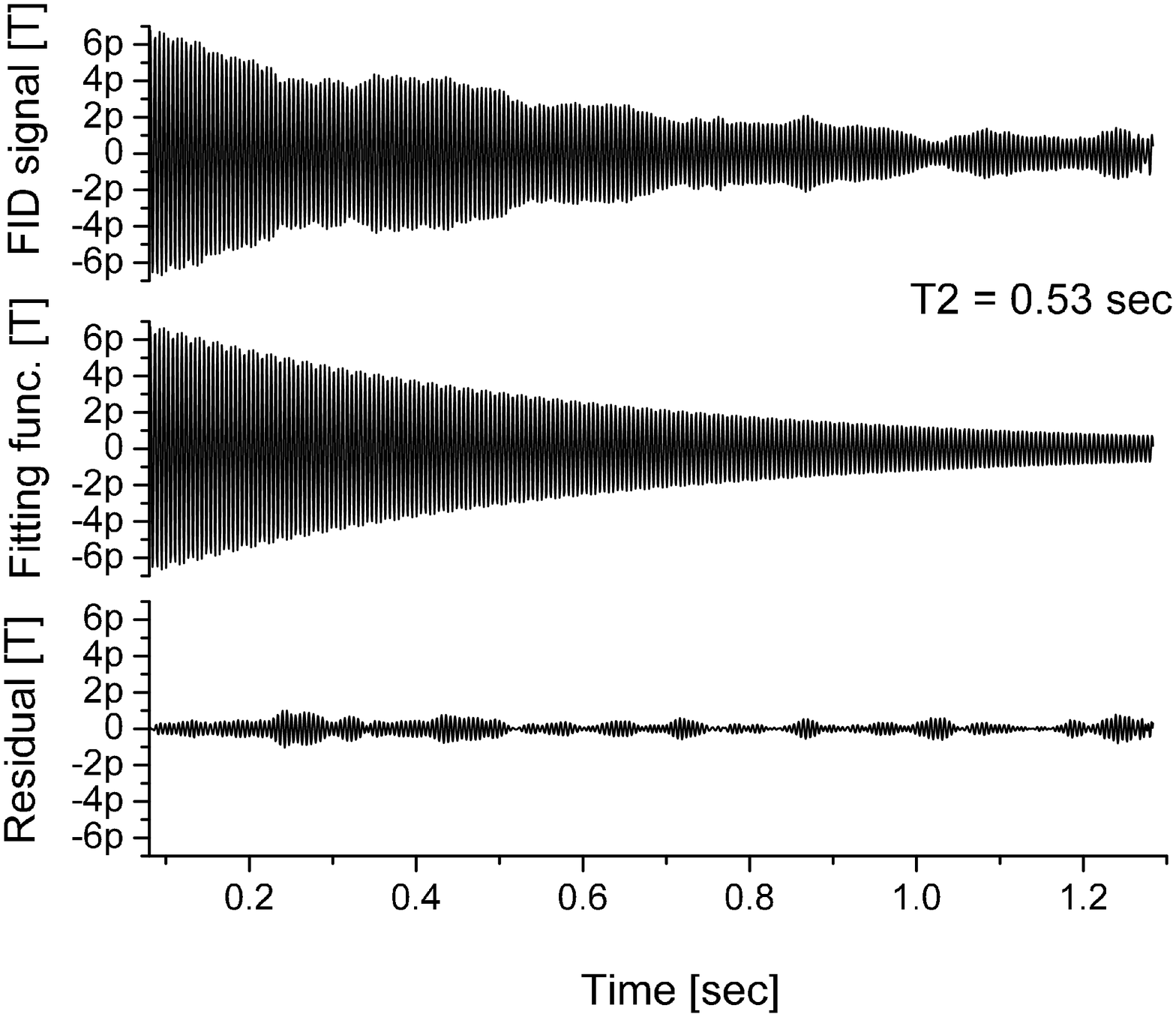}}
  \centerline{\includegraphics[width=7.5cm]{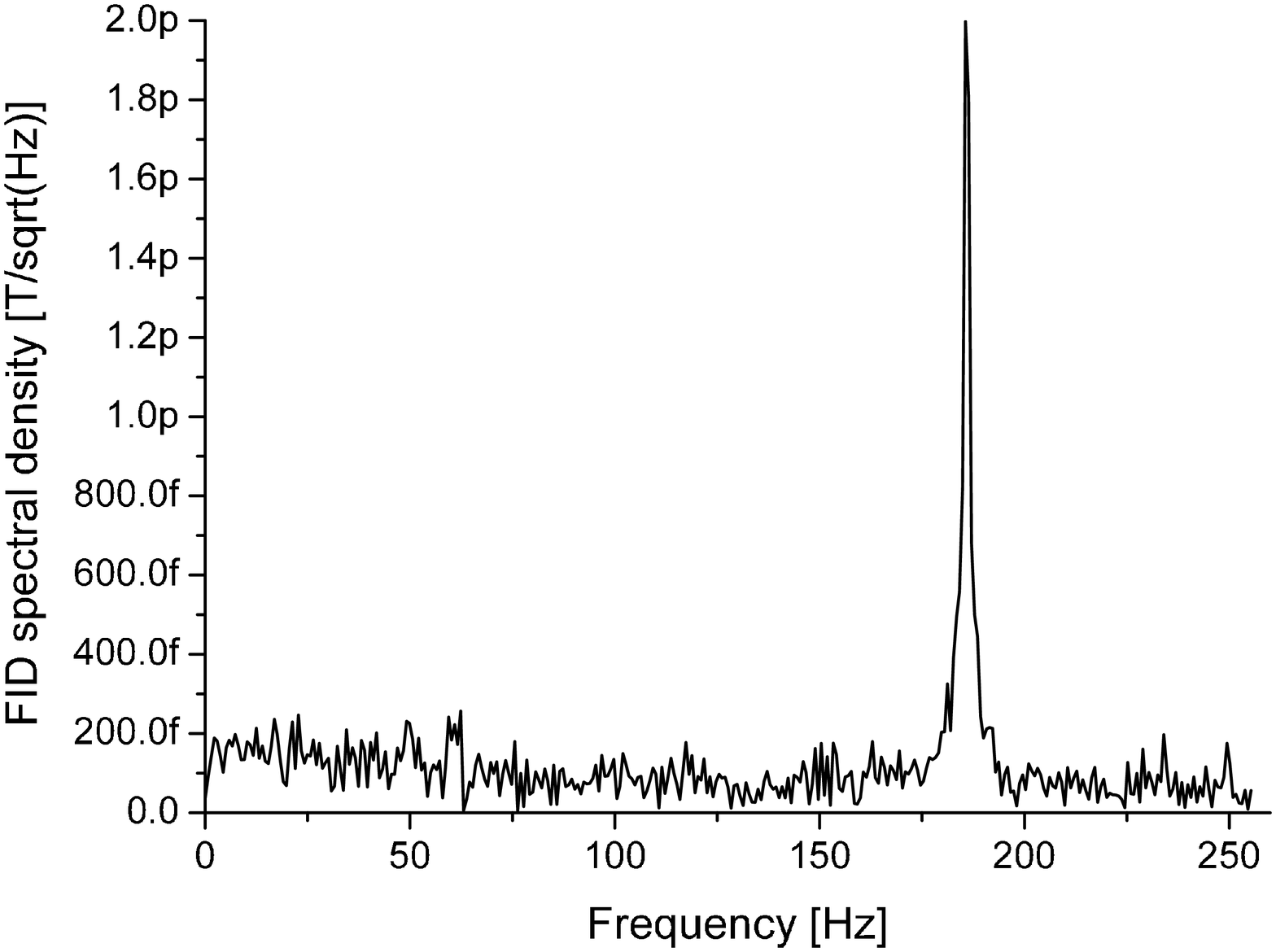}}
    \caption{Proton FID signal obtained by non-adiabatic rotation of the nuclear spins (time domain and power spectrum). The signal was obtained by averaging  1500 traces demodulated by the lock-in amplifier.  \label{nonadfid}}
\end{figure}

\begin{figure}
  \centerline{\includegraphics[width=7.5cm]{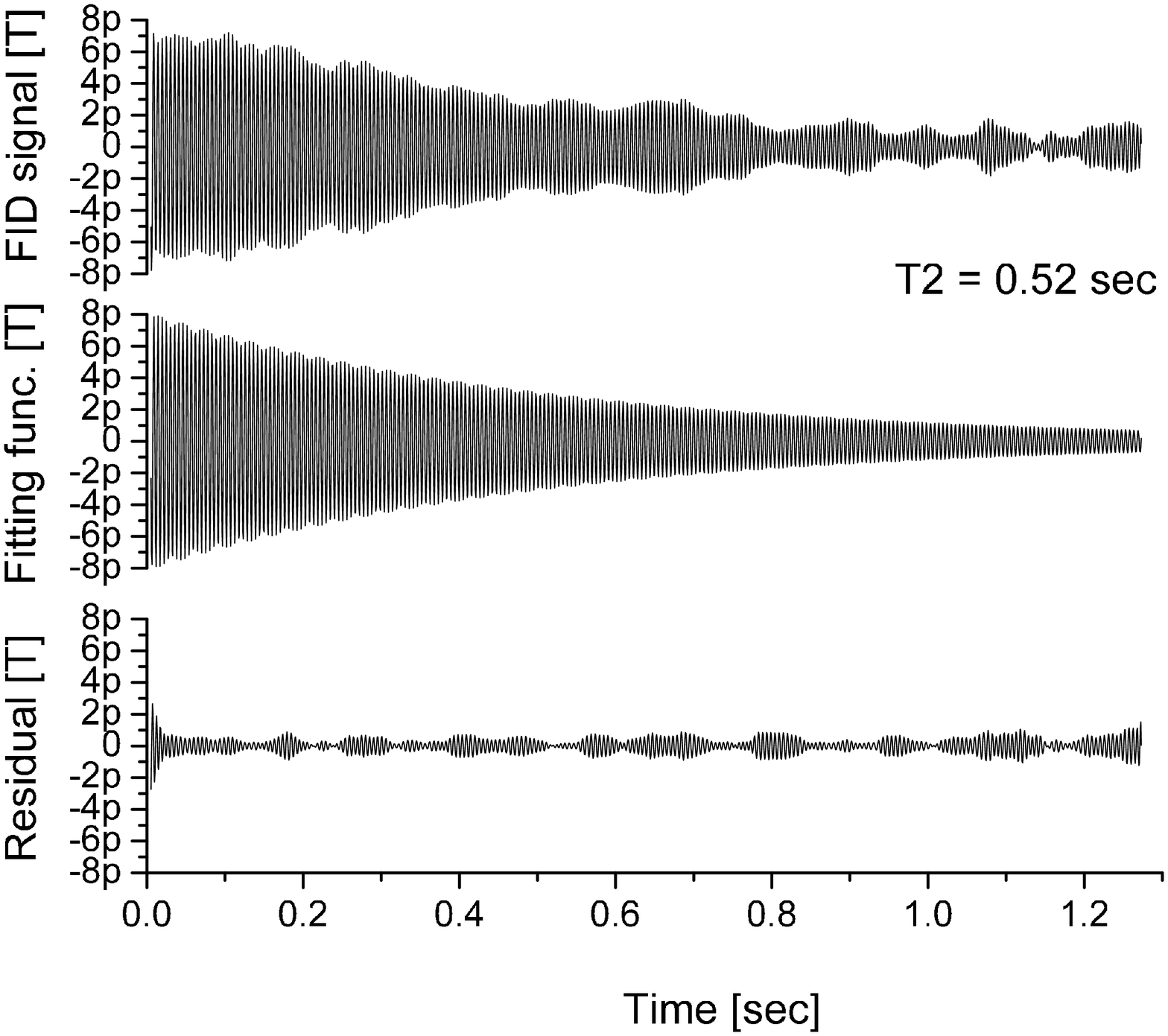}}
  \centerline{\includegraphics[width=7.5cm]{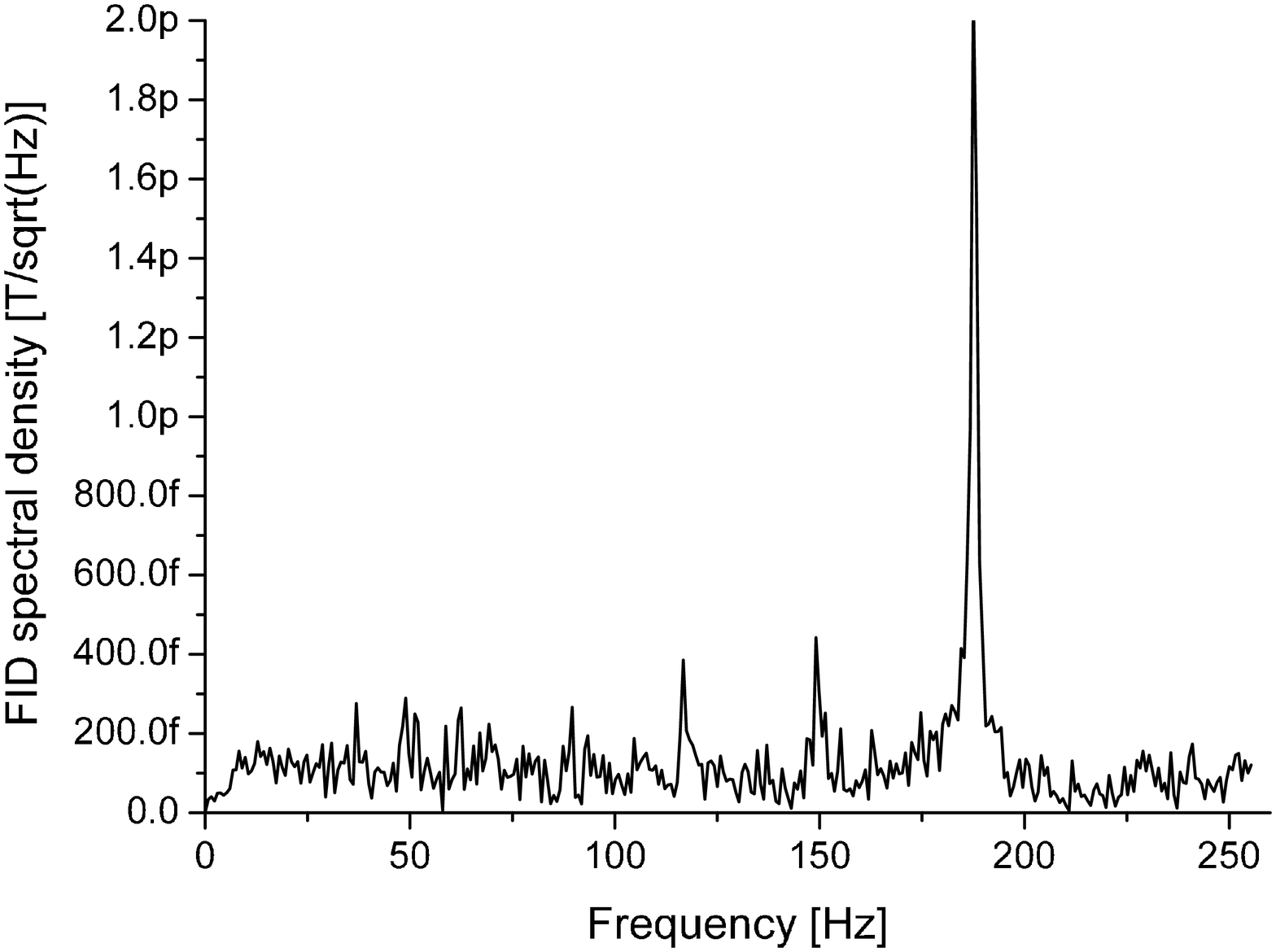}}
    \caption{Proton FID signal obtained by applying a $\pi/2$ pulse (time domain and power spectrum). The signal is obtained from the average of 730 traces demodulated by the lock-in amplifier. \label{pulsefid}}
\end{figure}

Fig. \ref{nonadfid}  shows the FID signal produced by non-adiabatic spin rotation and recorded by lock-in demodulation, as it appears in the time and  frequency domains. The time domain signal (upper trace) is shown together with the corresponding fitting curve (modelled as an exponentially decaying cosine,  $ A  \exp(-t/T_2) \cos(\omega t + \phi)$), and the residual. Similar results are obtained when manipulating the spin with ac pulses, as shown in Fig. \ref{pulsefid}.

\section{Discussion}
\label{discussion}

\subsection{Results and perspectives}
The good quality of the signal reported in the previous section, demonstrates that the low intrinsic  signal-to-noise ratio of LF-NMR in an unshielded environment can be overcome by using a low field only for the precession detection, while maintaining relatively high values for the polarization field. The latter, on the other hand, does not require a high level of homogeneity and (as in our case), can be  generated inexpensively by permanent magnets. Our set-up is not optimized for sensitivity, but  we address  some possible improvements to the sensor below, e.g. using more efficient pumping radiation and specially designed vapour cells, in order to maximize the sample-sensor coupling. In spite of the relatively low sensitivity, the stability of the system makes it capable of achieving quite high resolution, by means of automated, long-lasting measurements and averaging procedures for noise rejection.

\subsection{Detection limit}
\label{sensitivity}
The sensitivity can be accurately determined by measuring the noise level (which, following optimization, reaches the limit set by the shot noise in the photo-current of the polarimeter photo-diodes) and the slope of the atomic resonance used to make the system self-oscillate. Our apparatus permits fast  switching between forced and self-oscillating  operating modes, so the slope can be evaluated by measuring the amplitude of the polarimetric signal while scanning the frequency of the forcing signal around the atomic precession frequency. In fact, accurate evaluation of the slope is achieved by best-fit procedures. Repeated noise measurements and slope evaluation are the primary-best method for optimization of the magnetometric signal-to-noise ratio. As discussed in detail in \cite{belfijosa09}, several parameters need to be adjusted in order to achieve the optimal working conditions. To this end, we developed an automated procedure (see Appendix \ref{selfoptimize} for details) to identify this optimal working condition. As previously reported in \cite{belfijosa09}, our set-up works in stable conditions with a single arm sensitivity of 2~pT$/\sqrt{\mathrm {\mbox Hz}}$ and is mainly limited by a disadvantageous atomic transition, as discussed in Sec.\ref{nextsteps}.
The relatively low sensitivity (especially  compared to state-of-art low temperature SQUID and SERF magnetometers), is  partially compensated by several practical features. First of all, the system is very stable and can work autonomously for hours, making it possible to record hundreds or thousands of traces: the  recording time is usually only limited by the duration of the working day. The fact that the sensor head works at room
temperature (or just above), eliminates any constraints on the minimum distance between sample and sensor. This reduces  the effort necessary to minimize the head size, which, for small
size samples, leads to improve the sample-sensor coupling and thus the the NMR detection limit, even without  increasing the sensitivity. In fact, miniaturized magneto-optical sensors \cite{schwindtapl04} have already been developed, and excellent results have also been demonstrated in this field  \cite{ledbetterpnas08, ledbetter09}.

\subsection{Possible improvements of the system}
\label{nextsteps}

The atomic species and the specific atomic transition are important features that place both practical and fundamental limits on the performance of the OAM. The use of Cs is unfavorable compared to other alkali species because of its lower Land\'e factor and consequently lower gyromagnetic factor. On the other hand, the advantages of Cs include its higher vapour density at room temperature, the existence of a unique natural isotope, and very large  hyperfine separation of the ground states, resulting in a very weak non-linear Zeeman effect. A further advantage is given by the relatively large nuclear spin, which, leading to a large nuclear slow-down factor, causes a reduction of the broadening for spin-exchange collisions \cite{appelt98, seltzertesi08}.

As mentioned in Sec.\ref{atomicmagnetometer},  $D_2$ transitions are an unfavorable choice for optically pumping  alkali atoms. Nowadays, laser sources based on distributed feedback technology are available to excite the $D_1$ transition of Cs. This represents an important opportunity to improve the sensitivity of our apparatus by more than one order of magnitude. Optical pumping in the $D_2$ transition is weakly effective, as atoms in  all the ground state Zeeman sublevels may absorb the circularly polarized light due to the presence of an excited state with larger F number with respect to the ground state. The pumping effect is indirect under these conditions and in the presence of high pressure buffer gas  it results from a depletion mechanism: the circularly polarized light preferably excites large $m$ ground levels, due to larger Clebsch-Gordan coefficients, the populations  in the excited Zeeman sublevels are then equalized due to collisional mixing, leading  atoms to decay back to the ground sublevels with equal probability. Using $D_1$ transition is a much more efficient way to polarize atoms and is made even more efficient by the collisional mixing. In fact in $^2S_{1/2} \rightarrow ^2P_{1/2}$ transitions, atom in largest $m$ ground state cannot
absorb circularly polarized photons. In this case, the atomic sample is pumped directly and collisional mixing of the excited state makes the relative values of the Clebsch-Gordan coefficients irrelevant. They only play a minor role in enhancing  the pumping rate,  provided that the $\Delta m =0$ transitions are weaker than the $\Delta m
=1$ ones. See Chapter 2 in \cite{seltzertesi08} for a detailed discussion of these mechanisms and \cite{happer72} for a general review on optical pumping.

It is worth stressing that in accordance with Eq.(\ref{signalesp}) the amplitude of the peak corresponding
to the nuclear signal depends on $J_1(M)\approx M/2$, where M is the modulation index i.e. the ratio max$(\delta \omega_a)/\omega_n$ between the maximum deviation  of the instantaneous atomic precession frequency and the
nuclear precession frequency. The direct consequence of this  is that operating at higher precession fields, produces a proportionally weaker signal.  Similarly, Eq.\ref{lockinoutput} shows that the signal does not depend on the nuclear precession frequency  with the lock-in detection either. In conclusion, in contrast to conventional NMR, no advantages in terms of signal-to-noise ratio are obtained by increasing the precession field.

\section{Conclusions }
We have demonstrated that LF-NMR in the micro-Tesla range is feasible with an atomic magnetometer operating in an unshielded environment. We tested our apparatus in a remote detection experiment using both non-adiabatic and ac pulses to rotate nuclear spins with respect to the precession field. We considered two different approaches for demodulating the NMR signal from the polarimetric signal generated by the OAM, also taking advantage of the differential nature of our set-up. The relatively poor sensitivity could be significantly improved by increasing the optical pumping efficiency of the atomic sample, which would be possible using a laser source tuned to the $D_1$ line. With the current sensitivity level (which is about \pthz{2}), the long-term stability of the system also makes it possible to achieve high resolution via automated, long-lasting measurement and averaging procedures.

\appendix

\section{Description of the routine used for identifying the optimal OAM working point }
 \label{selfoptimize}
The working conditions of the self-oscillating magnetometer are controlled via a program interfacing with the waveform  generator that drives the diode laser frequency modulation. Having localized the resonance in scanned mode, the system is switched to self-oscillation regime by replacing the  asymmetric square-wave output with an indefinite train of single pulses triggered by the polarimetric signal of the main arm. The amplitude and width of the pulses are set on the basis of the optimal square-wave output previously identified in the scanned mode. The pulse delay is set to typical values and then optimized. An automated simplex-optimization procedure based on a simplex approach is then performed by adjusting four parameters of the pulse generator. The search for the optimal operating conditions is based on  maximization of the  polarimeter signal amplitude, which is evaluated by referencing the lock-in amplifier with the comparator output: the optimization procedure seeks the optimal conditions in a four-dimensional space by adjusting pulse delay,  width,  amplitude and dc offset. The latter parameter controls the average optical detuning of the pump laser, thus making it possible to counteract the slow drifts in the optical frequency that occur in
long-lasting measurements. We found that  the probe laser tuning is a less critical parameter, so that neither remote control nor the optimization procedure were included in the program for its value. A run of the optimization routine takes about 5\,s to 15\,s, mainly depending on the starting conditions and on the single-point measurement time, which is in turn set by the lock-in settling time.

\section*{Acknowledgments}
The authors are grateful to Emma Thorley of the {\it UniSi Translation Service} for reviewing the manuscript.


\begin{thebibliography}{99}
\footnotesize

\bibitem{longquingjmr08}
L.\,Qiu, Yi\,Zhang, H.-J.\,Krause, Alex I.\,Braginski,
A.\,Offenhäusser, 
Low-field NMR measurement procedure when SQUID detection is used, 
Journal of Magnetic Resonance { 196} (2009)  101-104.

\bibitem{greenbergrmp98} Ya.\,Greenberg, 
Application of superconducting quantum interference devices to nuclear magnetic resonance, 
Rev. Mod. Phys. { 70} (1998)  175-222.


\bibitem{mcdermottscience02} R.\,McDermott, A.\,H.\,Trabesinger, M.\,Mueck, E.\,L.\,Hahn, A.\,Pines, J.\,Clarke, 
Liquid-state NMR and scalar   couplings in microtesla magnetic fields, 
Science { 295}  (2002)  2247-2249.


\bibitem{mcdermottpnas04}
R.\,McDermott, S.\,K.\,Lee, B.\,Haken, A.\,H.\,Trabesinger, A.\,Pines, J.\,Clarke,
Microtesla MRI with a superconducting quantum interference device, 
Proc. National Acad. Sci. 101 (2004)  7857-7861.


\bibitem{romalisnature07}
D.\,Budker and M.\,V.\,Romalis, 
Optical magnetometry, 
Nature Physics { 3},  (2007), 227-234.

\bibitem{cohen69} C.\,Chohen-Tannoudji, J.\,Du\,Pont-Roc, S.\,Haroche, F.\,Laloe, 
Detection of the Static Magnetic Field  Produced by the Oriented Nuclei of Optically pumped $^3$He Gas, 
Phys. Rev. Lett. { 22}  (1969) 758-760.  


\bibitem{leeaapl06}
S.-K.\,Leea, K.\,L.\,Sauer, S.\,J.\,Seltzer, O.\,Alem, M.\,V.\,Romalis, Subfemtotesla radio-frequency atomic magnetometer for detection of nuclear quadrupole resonance, 
Appl. Phys. Lett. {89}  (2006) 214106.

\bibitem{kornackprl05}
T.\,W.\,Kornack, R.\,K.\,Ghosh, and M.\,V.\,Romalis, 
Nuclear Spin Gyroscope Based on an Atomic Comagnetometer, 
Phys. Rev. Lett. 95 (2005) 230801.

\bibitem{belfijosa07} J.\,Belfi, G.\,Bevilacqua, V.\,Biancalana, Y.\,Dancheva, L.\,Moi, 
All optical sensor for automated magnetometry based on coherent population trapping, 
J. Opt. Soc.  Am. { B 24}   (2007)  1482-1489.

\bibitem{matsko05} 
A.\,B.\,Matsko, D.\,Strekalov, L.\,Maleki, 
Magnetometer based on the opto-electronic microwave oscillator, 
Optics Communications { 247} (2005) 141-148.


\bibitem{weis03} G.\,Bison, R.\,Wynands, and A.\,Weis,
A laser-pumped magnetometer for the mapping of human cardiomagnetic fields,; 
Appl. Phys. B - Lasers Opt. { 76} (2003) 325-328.



\bibitem{savukovprl05a}
I.\,M.\,Savukov and M.\,V.\,Romalis, 
NMR Detection with an Atomic Magnetometer,
Phys. Rev. Lett. 94  (2005) 123001.

\bibitem{belfijosa09} J.\,Belfi, G.\,Bevilacqua, V.\,Biancalana, S.\,Cartaleva, Y.\,Dancheva, K.\,Khanbekyan, L.\,Moi, 
Dual channel self-oscillating optical magnetometer,
J. Opt. Soc.  Am. { B 26} (2009) 910-916.

\bibitem{savukovjmr07}
I.\,M.\,Savukov, S.\,J.\,Seltzer, M.\,V.\,Romalis, 
Detection of NMR signals with a radio-frequency atomic magnetometer,
Journal of Magnetic Resonance { 185}  (2007) 214-220.

\bibitem{ledbetter09} M.\,P.\,Ledbetter, C.\,W.\,Crawford, A.\,Pines, D.\,E.\,Wemmerb, S.\,Knappe, J.\,Kitching, D.\,Budker, Optical detection of NMR J-spectra at zero magnetic field, Journal of Magnetic Resonance { 199} (2009)  25-29.

\bibitem{savukovjmr09}I.\,M.\,Savukov, V.\,S.\,Zotev, P.\,L.\,Volegov, M.\,A.\,Espy, A.\,N.\,Matlashov, J.\,J.\,Gomez and R.\,H.\,Kraus Jr., MRI with an atomic magnetometer suitable for practical imaging applications, Journal of Magnetic Resonance { 199} (2009)  188-191

\bibitem{schwindtapl04} P.\,D.\,D.\, Schwindt, S.\,Knappe, V.\,Shah, L.\,Hollberg, J.\,Kitching, Li-A.\,Liew,  J.\,Moreland, Chip-scale atomic magnetometer, Appl.\,Phys.\,Lett. { 85} (2004) 6409-6411.

\bibitem{ledbetterpnas08}M.\,P.\,Ledbetter, I.\,M.\,Savukov, D.\,Budker, V.\,Shah, S.\,Knappe, J.\,Kitching, D.\,J.\,Michalak, S.\,Xu,  A.\,Pines, 
Zero-field remote detection of NMR with a microfabricated atomic magnetometer, 
Proc. National Acad. Sci. 105 (2008) 2286-2290.

\bibitem{seton05} H.\,Seton, J.\,M.\,S.\,Hutchison, D.\,M.\,Bussel, 
Liquid helium cryostat for SQUID-based MRI receivers,
Cryogenics, { 45}  (2005) 348-355.

\bibitem{kominisnature03} I.\,K.\,Kominis, T.\,W.\,Kornack, J.\,C.\,Allred,  M.\,V.\,Romalis, 
A subfemtotesla multichannel atomic magnetometer, 
Nature { 422} (2003) 596-599.

\bibitem{savukovpra05} I.\,M.\,Savukov, M.\,V.\,Romalis, 
Effects of spin-exchange collisions in a high-density alkali-metal vapour in low magnetic fields, 
Phys. Rev. { A 71}  (2005) 023405.

\bibitem{allredprl02} J.\,Allred, R.\,Lyman, T.\,Kornack, M.\,Romalis, 
A high-sensitivity atomic magnetometer unaffected by spin-exchange relaxation, 
Phys. Rev. Lett. { 89} (2002) 130801.

\bibitem{xursi06} S.\,Xu, S.\,M.\,Rochester, V.\,V.\,Yashchuk, M.\,H.\,Donaldson, D.\,Budker, 
Construction and applications of an atomic magnetic gradiometer based on nonlinear magneto-optical rotation, 
Rev. of Sc. Instr. { 77}  (2006) 083106.

\bibitem{savukovprl05b}
I.\,M.\,Savukov,  M.\,V.\,Romalis, 
Tunable Atomic Magnetometer for Detection of Radio-Frequency Magnetic Fields, 
Phys. Rev. Lett.  { 95}  (2005) 063004.

\bibitem{yangapl06}
H.\,C.\,Yang, S.\,H.\,Liao, H.\,E.\,Horng, S.\,L.\,Kuo, H.\,H.\,Chen, S.\,Y.\,Yang, 
Enhancement of nuclear magnetic resonance in microtesla magnetic field with prepolarization field detected with high-Tc superconducting quantum interference device,
Appl. Phys. Lett. { 88}  (2006) 252505.

\bibitem{byran08}
J.\,Byran,  A.\,Kantzas, A.\,Mai, Heavy Oil Reservoir Characterization Using Low Field NMR, contribution to CSPG CSEG CWLS
Convention (2008).

\bibitem{Aeberhardt06} K.\,Aeberhardt , Q.\,D.\,Bui, V.\,Normand, 
Using low-field NMR to infer the physical properties of glassy oligosaccharide/water mixtures, 
Biomacromolecules { 8}  (2007) 1038-1046. 

\bibitem{andrade06}
L.\,Andrade, I.\,Farhat, 
An investigation into the practical aspects of the quantitative analysis of solid-liquid systems using low-field NMR, 
Contribution to The 8th International Conference on the Applications of Magnetic Resonance in Food Science, 2006, Nottingham.

\bibitem{callaghanajp97}
P.\,T.\,Callaghan, C.\,D.Eccles, J.\,D.\,Seymour, 
An Earth's field NMR apparatus suitable for Pulsed Gradient Spin Echo measurements of self-diffusion under Antarctic conditions, 
Rev. Sci Instr. { 68} (1997) 4263-4270;

P.\,T.\,Callaghan, C.\,D.\,Eccles, 
NMR studies of Antarctic Sea Ice,
Bulletin of Magnetic Resonance { 18}  (1996) 62-64;

P.\,T.\,Callaghan and M.\,Le Gros,
 Nuclear spins in the earth's magnetic field, 
American Journal of Physics {50} (1982) 709-713.


\bibitem{belfijosacardio07} J.\,Belfi, G.\,Bevilacqua, V.\,Biancalana, S.\,Cartaleva, Y.\,Dancheva, L.\,Moi; Cesium coherent population trapping magnetometer for cardiosignal detection in an unshielded environment,
J. Opt. Soc.  Am. { B 24}  (2007) 2357-2362.

\bibitem {lanczosgrandke} C.\,Lanczos, Applied Analysis, Dover Publications Inc. (NY 1988), originally published by Englewood Cliffs, N.J. Prentice-Hall (1956), par. III.5, IV.22;

T.\,Grandke,  
Interpolation algorithms for discrete Fourier transforms of  weighted signals, 
IEEE Trans. Instrum. Meas. IM-32 (1983) 350-355.


\bibitem{kau01}
J.\,Kauppinen, J.\,Partanen, 
Fourier Transform in Spectroscopy, 
Chap.11, Ed. Wiley-VCH, Berlin (2001).

\bibitem{appelt98} S.\,Appelt, A.\,Ben-Amar Baranga, C.\,J.\,Erickson, M.\,V.\,Romalis, A.\,R.\,Young, and W.\,Happer (1998). Theory of spin-exchange optical pumping of 3He and 129Xe. Phys. Rev. A 58 (2), 1412–1439.

\bibitem{seltzertesi08} S.\,J.\,Seltzer, 
Developments in Alkali-Metal Atomic Magnetometry, 
Dissertation, Princeton University (2008). \newline
Available at http://physics.princeton.edu/romalis/papers/Seltzer Thesis.pdf

\bibitem{happer72} W.\, Happer, Optical Pumping, Rev. Mod. Phys 44 (1972) 169-250


\end{thebibliography}
\end{document}